\documentclass[journal]{vgtc}                     

\onlineid{1405}

\vgtccategory{Research}

\vgtcpapertype{please specify}

\title{The Impact of Navigation on Proxemics in an Immersive Virtual Environment with Conversational Agents}

\author{%
  \authororcid{Rose\ Connolly}{0009-0009-6531-2973} 
  \authororcid{Lauren\ Buck}{0000-0002-7220-3558}
  \authororcid{Victor\ Zordan}{0000-0002-7309-7013}
  \authororcid{Rachel\ McDonnell}{0000-0002-1957-2506}
}

\authorfooter{
  \item
  	Rose Connolly \& Rachel McDonnell are with Trinity College Dublin.\break
  	E-mail: connolr3@tcd.ie\break
    Email: ramcdonn@tcd.ie
  \item Lauren Buck is with University of Utah.
  Email: lauren.buck@utah.edu
  \item
  	Victor\ Zordan is with Roblox Corp.
  	E-mail: vbzordan@roblox.com

}

\abstract{
As social VR grows in popularity, understanding how to optimise interactions becomes increasingly important. Interpersonal distance—the physical space people maintain between each other—is a key aspect of user experience. Previous work in psychology has shown that breaches of personal space cause stress and discomfort. Thus, effectively managing this distance is crucial in social VR, where social interactions are frequent. Teleportation, a commonly used locomotion method in these environments, involves distinct cognitive processes and requires users to rely on their ability to estimate distance. Despite its widespread use, the effect of teleportation on proximity remains unexplored. To investigate this, we measured the interpersonal distance of 70 participants during interactions with embodied conversational agents, comparing teleportation to natural walking. Our findings revealed that participants maintained closer proximity from the agents during teleportation. Female participants kept greater distances from the agents than male participants, and natural walking was associated with higher agency and body ownership, though co-presence remained unchanged. We propose that differences in spatial perception and spatial cognitive load contribute to reduced interpersonal distance with teleportation. These findings emphasise that proximity should be a key consideration when selecting locomotion methods in social VR, highlighting the need for further research on how locomotion impacts spatial perception and social dynamics in virtual environments.

} 


\teaser{
  \centering
  \includegraphics[width=\linewidth]{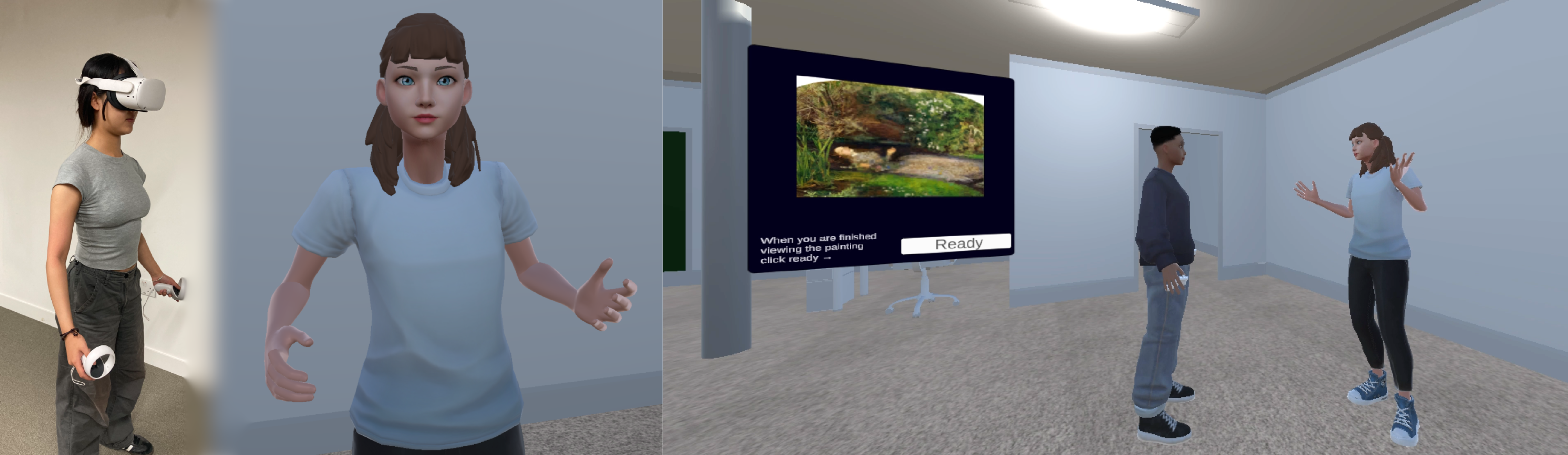}
  \caption{(left to right), participant with HMD and controllers immersed in the virtual scene, the participant's first-person view of the conversational agent, and a third-person view of the scene where the participant is talking with the conversational agent on the right. They are discussing details about the painting `Ophelia' which is shown in background for illustration purposes. In the experiment, participants either walked naturally or teleported to the agent before beginning the discussion.
  }
  \label{fig:rpm experiment girl}
  }

\graphicspath{{figs/}{figures/}{pictures/}{images/}{./}} 

\usepackage{float}

\usepackage{mathptmx}                  

\usepackage[normalem]{ulem}

\usepackage{mathptmx}                  
\usepackage{array}
\usepackage{multirow}
\usepackage{float}
\usepackage{tabularx}  
\usepackage{graphicx}  
\usepackage{ragged2e}

\usepackage{booktabs}                  
\usepackage{lipsum}                    
\usepackage{times}                     
\usepackage{enumitem}
\usepackage{mathptmx}                  

\begin{document}


\firstsection{Introduction}

\maketitle
The study of proxemics, which explores how individuals perceive and use space in social interactions \cite{hall1966hidden}, has emerged as a significant area of research in the virtual reality (VR) community. In proxemics, interpersonal distance (IPD) - the physical space people keep between each other — reflects social dynamics, as individuals adjust their spatial behavior based on their relationships and environment. The violation of interpersonal space is deeply uncomfortable and stressing for users, provoking a number of negative responses \cite{kanaga1981relationship, welsch2019anisotropy}. As such, the designers of social VR environments should consider proximity in order to prevent the elicitation of negative feelings in users.

The choice of locomotion method, such as teleportation or natural walking, plays a crucial role in shaping user experience, as each has distinct spatial perception considerations \cite{cherep2020spatial, schneider2018locomotion}. Teleportation, commonly used in social VR settings for navigating large environments, allows users to instantly ``jump'' to new locations, bypassing physical space constraints. However, teleportation also poses challenges related to spatial orientation \cite{cherep2020spatial, bowman1997travel, bakker2003effects} and has greater spatial cognitive costs compared to natural walking \cite{cherep2020spatial}. Unlike walking, teleportation lacks proprioceptive and vestibular feedback, which are important for accurate distance estimation \cite{campos2014contributions, keil2021effects}. This disruption in spatial navigation -- caused by the absence of continuous visual movement -- makes it harder for users to adjust to new spatial cues and maintain orientation \cite{bhandari2018teleportation, paris2019video}. 

In contrast, natural walking provides continuous self-motion cues, and spatial updating and navigation are enhanced with this method \cite{kelly2021effectiveness}. However, concerns about potential collisions with real-world objects and limitations in physical space may arise. Walking provides both vestibular and proprioceptive feedback that is also associated with increased embodiment \cite{leonardis2014multisensory}. This suggests continuous locomotion methods have higher levels of agency and body ownership (aspects of embodiment \cite{kilteni2012sense}), which we examine in our study. Furthermore, continuous locomotion methods have been found to enhance co-presence compared to teleportation, in the context of artificial locomotion \cite{freiwald2021effects}. Since the impact of natural walking on co-presence relative to teleportation has not been thoroughly explored, we aim to investigate whether this effect also applies to natural walking. Given the distinct effects of natural walking and teleportation on spatial processes, we hypothesised that IPD, as a mediator of perceived distance, will be affected.

In our study, we conducted an experiment where participants interacted with female embodied conversational agents (ECAs), which are virtual humans capable of both verbal and non-verbal communication \cite{cassell2001embodied}. Participants used either teleportation or natural walking during these interactions, and we evaluated their proximity to the agents, perceived agency, body ownership, co-presence, and overall user experience through trials and questionnaires.

Prior research suggests that the locomotion method employed in an immersive virtual environment can influence distance perception, with different locomotion methods producing different distance estimation biases~\cite{keil2021effects,maruhn2019measuring}. Building on this, our study is the first to investigate how the type of locomotion method influences interpersonal distance between users in an immersive virtual environment. 

Our main finding, that users tend to get closer to virtual agents when teleporting, could influence perceptions of personal boundaries and social norms in virtual environments. This may lead to interactions becoming more informal or intense, as the conventional sense of personal space is disrupted. VR developers may need to adjust interaction distances to strike a balance between immersion and comfort or implement features such as virtual personal boundaries to help regulate proximity and protect users' personal space.

\section {Background and Related Work}

\subsection{Proxemics}

Proxemics, coined by Edward Hall~\cite{hall1966hidden}, is the study of how humans use the space around the body during social interaction. Hall divided this space into four distinct zones: intimate ($<0.45m$), personal ($0.45m-1.2m$), social ($1.2m-3.6m$), and public ($>3.6m$). The regulation of personal space, or the space around one's body that is considered one's own, is one component of the study of proxemics. Different social factors influence how an individual regulates interactions within this space, and when this space is intruded upon, one can experience discomfort, anger, or anxiety~\cite{hayduk1983personal}.

VR presents a unique opportunity to study proxemics, as it enables precise 
measurements of space in highly controlled environments that are have the potential to facilitate naturalistic interaction. Proxemics are often measured in these environments via the interpersonal distance (IPD) that users keep between one another~\cite{buck2020determining,llobera2010proxemics,miller2023large,zibrek2020effect}. Such studies involve particular modes of interaction, with either the participant approaching (active interaction)~\cite{iachini2014body} or being approached (passive interaction) by an interactant~\cite{buck2020determining}. The mode of interaction has been shown to affect IPD, as IPD increases during passive compared to active interaction \cite{iachini2014body}. 

Other studies document the effect of age and gender on proxemics. Female participants have been found to maintain larger distances from virtual humans compared to males~\cite{zibrek2020effect,iachini2014body}, while male participants keep shorter distances between themselves and and female agents~\cite{iachini2016peripersonal}. While these findings highlight consistent trends, gender effects remain unclear in certain scenarios. Proximity varies across many contexts including the method of interaction between agent and participant gender \cite{iachini2016peripersonal}, as well as the nature of the interactant (object, avatar, confederate, etc.) \cite{iachini2014body}. These are among the often overlapping influences on proximity that prevent definitive conclusions from being formed.

As noted by Welsch et al. \cite{welsch2021proxemics}, the understanding of proximity in social VR remains incomplete. Many VR studies on proxemics involve simulated situations with dyads, often lacking conversation (including  \cite{iachini2016peripersonal, zibrek2020effect, white1975interpersonal}), potentially rendering these results  ungeneralizeable to more realistic scenarios. 
Many components of social interaction such as verbal communication are often excluded from virtual environments used in proxemics studies. Recent studies have started to explore proximity in such contexts, such as group discussions \cite{miller2023large} and VR workshops \cite{williamson2021proxemics}, which better reflect more realistic scenarios in VR.

As IPD is frequently assessed in VR (examples include \cite{bailenson2003interpersonal, miller2023large, kim2024new, zibrek2020effect, iachini2016peripersonal}), it is important to understand any distinct effects specific to this medium. In general, the observed effects on distance in VR closely resemble those found in the real world. For example, the size of the room has been shown to be inversely related to IPD in both real \cite{white1975interpersonal} and virtual rooms \cite{miller2023large}. 
While the general patterns of IPD are mirrored in VR, one study suggests that these distances are exaggerated in VR \cite{kim2024new} i.e., people would stand further from each other in VR than in real life. As VR is inherently different from real life, distinct effects on proximity are to be expected. One area of interest in our study is how users traverse virtual environments. The methods of locomotion in VR, such as teleportation, differ significantly from real-life movement and may have distinct effects on IPD and social interactions.

\subsection{Distance Estimation}

A crucial element of spatial perception is one's ability to estimate distance, which can be described from different perspectives: egocentric (relative to the observer), exocentric (relative to another object), or allocentric (relative to a fixed location) \cite{sharika2014use, klatzky1998allocentric}. As outlined in the literature (see El Jamiy and Marsh \cite{el2019distance} and Renner et al.~\cite{renner2013perception} for in-depth reviews), distance is consistently estimated incorrectly in VR. Since our work focuses on the perception of egocentric and exocentric distance, this is where we focus our literature review. Egocentric distances are consistently underestimated in immersive virtual environments~\cite{bach2022effect,saracini2020differences}, but there are mixed results regarding exocentric distances~\cite{lin2014distance,waller1999factors,wartenberg2003precision}. 

There are several factors found to affect the estimation of egocentric distances in VR. For example, Saracini et al.~\cite{saracini2020differences} noted males and females to estimate egocentric distances with different levels of precision in real and virtual environments. The scale of a virtual room can also be associated with the degree of underestimation~\cite{bach2022effect}, as well as the FOV and weight of the head-mounted display~\cite{buck2018comparison,waller1999factors}, along with stereoscopic depth cues~\cite{wartenberg2003precision}. 

Geuss et al.~\cite{geuss2012effect} also suggest the planes of measurement has an effect on distance estimates in VR. This study compared VR to real life estimates, finding underestimations for egocentric distances and accurate estimations for exocentric distances. Distance estimates were also compared across the sagittal (depth) and frontal plane (width), and it was found that underestimation occurs for both distances in the depth plane (VR compared to real life), but no discrepancy occurred in distances along the frontal plane. Geuss et al.~\cite{geuss2012effect} highlights this is line with distance estimation studies conducted in the real world, showing an underestimation in the depth plane compared to the width plane \cite{kudoh2005dissociation, loomis1992visual}. 

In general research suggests that estimation errors in VR increase as distance increases~\cite{gagnon2020far,ng2016depth}. Distance estimation errors in peripersonal space (space immediately surrounding an individual) are generally small in VR \cite{armbruster2008depth, naceri2011depth} and in some contexts overestimated \cite{armbruster2008depth}. Errors in extrapersonal space (space beyond the immediate surroundings) seem to be more pronounced and subject to underestimation \cite{armbruster2008depth,keil2021effects, ng2016depth}. However, some studies present contrasting results. Saracini et al. \cite{saracini2020differences} found significant underestimations in peripersonal space but not in extrapersonal space when comparing VR to real life. They attribute this discrepancy to a lack of depth information from closer objects. Additionally, there are inconsistencies regarding the distance at which overestimations begin; Naceri et al. \cite{naceri2011depth} observed overestimations starting at 55 cm, while Armbruster et al. \cite{armbruster2008depth} noted this effect at 100 cm. Overall, existing research suggests that people are generally more accurate at estimating distances and sizes of objects that are close to them compared to those that are farther away. Nonetheless, the variability in findings indicate that further research is needed to fully understand distance estimation biases in both near and far spaces.

\subsection{Locomotion}
The locomotion method chosen by a developer impacts user experience in immersive virtual environments; it impacts factors such as immersion and comfort~\cite{lochner2021vr, boletsis2019vr}. Various locomotion techniques have been extensively researched, with common methods including natural walking, walking-in-place, teleportation, and continuous locomotion.

Walking is generally considered to provide a more immersive experience than teleportation \cite{lochner2021vr}. Studies consistently show that users often prefer walking, citing a heightened sense of presence \cite{sayyad2020walking, slater1995taking}. Walking is a continuous locomotion, so user's traversal of the environment happens fluidly and without visual interruption. Continuous navigation techniques have been found to increase co-presence, although this aspect is less frequently studied in relation to locomotion methods \cite{freiwald2021effects}.

Compared to teleportation, walking also offers varying degrees of embodiment, which refers to the extent to which users feel a virtual body belongs to them \cite{kilteni2012sense}.  The sense of embodiment involves three main components: agency, self-location, and body ownership \cite{kilteni2012sense}. Walking provides rich vestibular and proprioceptive cues that contribute to a stronger sense of embodiment 
\cite{leonardis2014multisensory}, and incorporating these cues artificially can even create illusionary embodiment. Research comparing continuous and non-continuous locomotion techniques indicates that embodiment tends to be stronger with continuous methods. However, studies comparing different continuous locomotion methods such as walking, walking-in-place, and continuous steering have found no significant differences among them \cite{dewez2020studying}. This suggests the continuity of locomotion is significant for user embodiment (whereas the specifics of continuous implementation are not).

In contrast to walking, teleportation eliminates physical space constraints. In controller-based teleportation, users must use their hands to select a destination and activate teleportation. The interaction effort originates in the hands, rather than the legs, which can lead to arm fatigue \cite{lou2020empirical}. However some studies suggest that overloading the hands with navigation functionality is not cognitively taxing. A study by Griffin et al. \cite{griffin2018evaluation} found that teleportation had a lower cognitive load compared to non-controller methods, although this was in comparison to head-tilt and walking in place rather than natural walking. Similarly another study \cite{loup2019effects} found a decreased cognitive load of teleportation, this time in comparison to arm swinging. However others argue that this navigation method should increase cognitive load, especially since users in many VR environments also need to use controllers for other types of interactions \cite{laviola2001hands} \cite{griffin2018evaluation}. To mitigate such issues, non-controller-based methods, such as gaze-based teleportation \cite{linn2017gaze} and teleportation activated by wink or mouth gestures \cite{prithul2022evaluation}, have been proposed. In terms of spatial cognitive processes, teleportation seems to be negatively impact cognitive map-building and overall ability to form spatial knowledge of the VE \cite{sayyad2020walking} \cite{langbehn2018evaluation}. 

Since teleportation does not generate optical flow, it is effective at reducing simulation sickness \cite{bowman1997travel}. However, because users do not visually traverse the space between their starting point and destination, they miss out on self-motion cues. This can result in spatial disorientation \cite{cherep2020spatial}. Additionally, teleportation makes it difficult for users to accurately estimate the distance traveled, often leading to underestimation \cite{keil2021effects}. As noted by \cite{campos2012multisensory}, this discrepancy can occur in locomotion methods that lack proprioceptive and vestibular cues, which typically aid in distance estimation. Such sensory feedback has also been shown to enhance embodiment \cite{leonardis2014multisensory}, underscoring the importance of choosing an appropriate locomotion method.

To address spatial disorientation in teleportation, non-continuous scene transitions like ``dashing" have been proposed \cite{griffin2018evaluation}. Additionally, teleportation methods with built-in rotation components have been developed to help users maintain spatial orientation \cite{wolf2021augmenting}. However including an integrated  rotational component may deter users from self-rotating which has been argued to be important for path integration \cite{lim2020rotational}. 

Many locomotion studies, (including both of the above adapted methods) evaluate with object based tasks. However, in this study we propose that the spatial disorientation common with teleportation may also impact social interactions, particularly regarding IPD mediation which depends on accurate distance estimation.

\subsection{Embodied Conversational Agents}
As stated, VR provides a controlled medium to facilitate proximity measurements. To represent both the interactant and the user when measuring proximity, virtual humans (VHs) are commonly used. VHs encompass both avatars, which are controlled by specific users, and agents, whose actions are determined by computer algorithms \cite{von2010doesn}. 

Of specific interest to this study is social interactions in VR. This can be facilitated through the use of embodied conversational agents (ECAs). In VR ECAs are represented by a virtual body often displaying gestures and movement (embodied). They interact with users by engaging in dialogue (conversational), determined by algorithms designed to simulate human like behaviour (agents) \cite{cassell2001embodied}.

Several factors contribute to the effectiveness of ECAs in VR, including their non-verbal communication. One study \cite{pejsa2017me} found that through non-verbal behaviours virtual agents have the capability to regulate turn-taking and indicate conversational roles.  This capability to influence user behavior is an interesting interaction dynamic, and this effect was uniquely observed in VR (compared to desktop settings) \cite{pejsa2017me}. Another study demonstrated higher levels of social presence when using embodied agents in VR contexts, once again in comparison to desktop \cite{guimaraes2020impact}. Together, these studies suggest that the use of embodied agents is particularly effective in VR due to its immersive nature.

Many other studies further explore the factors contributing to the effectiveness of ECAs in VR. Bailenson et al. \cite{bailenson2005independent} emphasise that consistency in matching the realism of an agent's appearance and behavior is important for co-presence, highlighting the importance of design fidelity. Another key consideration is the Uncanny Valley effect, first introduced by Mori \cite{mori1970bukimi}, which arises from subtle imperfections in human-like agents. Thaler et al. \cite{thaler2020agent} found that increased human-likeness in an agent's appearance increases perceptions of eeriness, though this effect varies across different ECA features. For example, Ferstl et al. \cite{ferstl2021human} observed that users prefer optimal realism in motion and voice, while Zibrek et al. \cite{zibrek2018effect} emphasized the role of appearance and personality in shaping affinity for virtual humans.

Despite these findings, the impact of uncanniness on proximity remains unclear. For instance, Zibrek et al. \cite{zibrek2018effect} reported no significant effect of render style on proximity, except in cases of closer proximity to a zombie character, which was attributed to curiosity rather than comfort. Similarly, earlier work by Zibrek et al. \cite{zibrek2017don} found no notable effect of realism on proximity, again looking at render style. However research is yet to establish the full effect of uncanniness on proximity behaviors - realism extends beyond render style to include dimensions such as motion and voice. The full influence on proximity remains unclear. Future studies are needed to clarify these relationships and investigate how various aspects of realism shape both affinity and spatial interactions.

Other behavioural findings have emerged regarding the emotional state of avatars. A study by Kim et al. \cite{kim2024engaged} found that agents appearing engaged in tasks are preferred over those showing strong emotions, which improves trust and presence.  Research concerning ECA also discusses the use of AI in creating features such as autonomy and responsive emotions that improve how agents interact \cite{luck2000applying}. In general lots of research concerning both agents and avatars is examined in contexts of training, education or health \cite{kyrlitsias2022social}, rather than for purely social interactions.

Overall, communication with ECAs are crucial for the effectiveness of many VR applications, but more research is required to fully understand how to optimise these interactions \cite{kyrlitsias2022social}. As highlighted, the importance of ECA stems from their ability to impact aspects of the user experience such as social presence and trustworthiness.  Through various features such as gestures and eye gaze, ECAs are effective ways to provide social interaction, augmented by the immersion of VR.

\section {Objectives and Hypotheses}
The objective of our study is to explore how proximity is influenced by different locomotion methods, specifically comparing teleportation to natural walking. These locomotion methods involve distinct spatial cognitive processes and tendencies toward overestimating or underestimating distances. We propose that these differences will impact IPD, which relies on distance estimation. This leads us to our primary hypothesis: 
 
\begin{itemize}
    \item[] \textbf{H1:} Participant's IPD to the agent will differ when participants approach using teleportation compared to walking.           
\end{itemize}

Due to inconclusive research and uncertainty about the specific cognitive processes involved in teleportation, we do not predict the exact nature of this difference. However, the unique cognitive demands \cite{cherep2020spatial} and spatial challenges \cite{bhandari2018teleportation, paris2019video}, associated with teleportation are expected to impact user-agent proximity. Teleportation lacks the self-motion cues of natural walking and presents users with different perspectives and measurement planes, all of which have been shown to introduce biases in distance estimation \cite{campos2014contributions,lin2014distance,geuss2012effect}. Since IPD involves the moderation of perceived space, we anticipate that teleportation, with its distinct spatial cues and varying estimation mechanisms, will influence IPD differently than natural walking.

We are also interested in the effect of locomotion method on agency, body ownership, and co-presence, along with potential gender-related differences. This leads us to formulate three secondary hypotheses:

\begin{itemize}

  \item[] \textbf{H2:} Body Ownership and agency will be greater for walking.

\end{itemize}

Previous research has shown that vestibular and proprioceptive cues, present in walking but not in teleportation, enhance embodiment \cite{leonardis2014multisensory}, which includes body ownership and agency \cite{kilteni2012sense}.

\begin{itemize}
    
    \item[] \textbf{H3:} Co-presence will be greater for walking.

\end{itemize}
    
Previous research \cite{freiwald2021effects} suggests that continuous locomotion may yield higher co-presence.

\begin{itemize}
  \item[] \textbf{H4:} Female participants will maintain greater proximity distances from the embodied agents than male participants. 
\end{itemize}

Previous studies have shown female participants keep greater interpersonal distances than male participants \cite{iachini2016peripersonal,zibrek2020effect}. We predict the same gender difference for this experiment. Agent gender have also been shown to have effects on interpersonal distance \cite{bailenson2003interpersonal}. Therefore we kept the agent's gender constant (always female) throughout the study in order to isolate the effects of locomotion and participant gender.

\section{Experimental Setup}

\begin{figure}[!bh]
    \centering
    \includegraphics[scale=0.3]{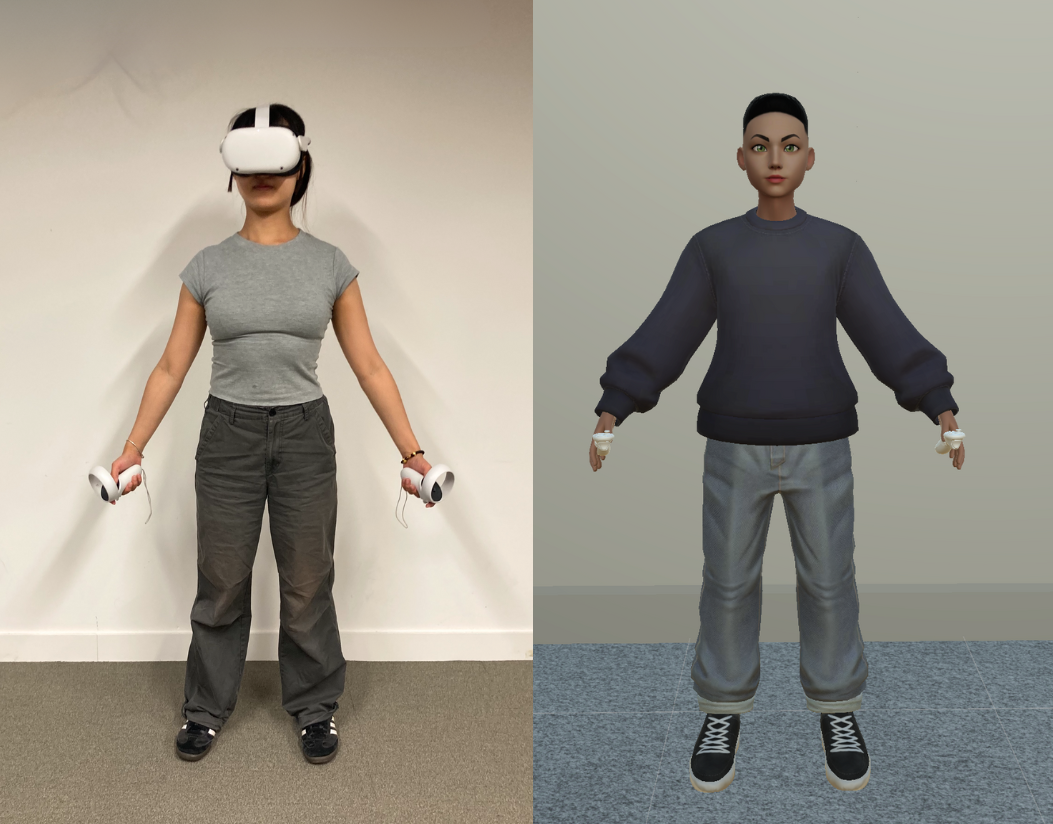}
    \caption{User (left) and avatar they were embodied in (right). Note that user only saw themselves in first-person - there was no reflective surface.}
    \label{fig:embodyavatar}
\end{figure}

Participants used the Oculus Quest 2 Head-Mounted Display (HMD) for the experiment, which was developed in Unity 2022.3.15.
The virtual environment was a plain environment with minimal clutter but depth cues such as a door and computer desk and chair (see \cref{fig:rpm experiment girl}). The virtual room was the same dimensions (\textit{10m X 6.5m}) as the room where the experiment took place such that the participant could walk around. Participants embodied an avatar from Ready Player Me, designed with a gender-neutral appearance. The avatar wore plain, baggy attire to hide their body shape and was styled in a black jumper and blue jeans, shown in \cref{fig:embodyavatar}. The virtual body’s height was set to be the same as the participant’s height using the floor as the tracking origin.

\subsection{Body Tracking \& Locomotion Implementation}
For body-tracking, we used MetaMovement SDK\footnote{https://developer.oculus.com/documentation/unity/move-overview/} which tracks the traditional 3-point of the hands and headset and uses AI to animate plausible leg movements based on these points, known as `generative legs'.

\begin{figure}[ht]
    \centering
    \includegraphics[scale=0.15]{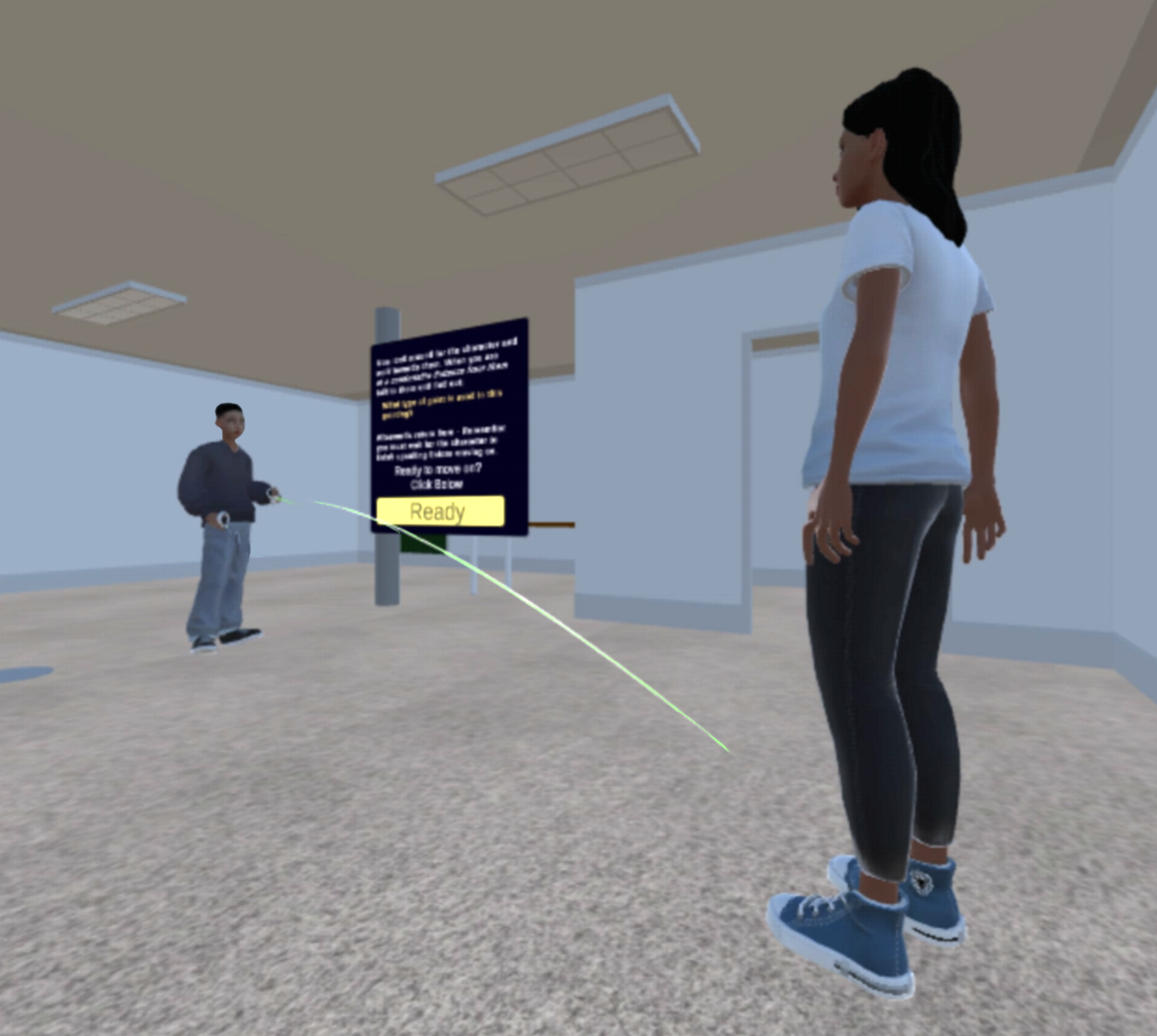}
    \caption{Third-person view of a participant (left) teleporting towards an embodied conversational agent (right).}
    \label{fig:teleporting}
\end{figure}

Teleportation was implemented with a visible trajectory ray that was curved. There was no reticle around the target destination, see \cref{fig:teleporting}. Participants could teleport with either controller. There was no rotation component available to participants in either locomotion method, instead participants were told they could turn in place. There were no limits to the number of teleports participants could make.  

\subsection {Embodied Conversational Agent}

We aimed to create conversational agents that would be knowledgeable about art, could respond naturally without any pre-scripted dialogue, and were able to interpret the participants speech well. We used Inworld AI's\footnote{https://www.inworld.ai/} SDK for Unity (\textit{AI NPC Engine - Dialogue \& Behavior for Unity - Inworld - Version 3.1.1.}) which enables the development of customizable, AI-driven characters powered by large language models (LLMs). Inworld AI uses GPT-3 as one of 20 machine learning models to provide basic natural language generation. Inworld's own frameworks builds layers of complexity upon this: including character memory, background and personality. The platform also provides third party integration with Ready Player Me\footnote{https://readyplayer.me/} so we could create a variety of virtual human appearances.

To ensure realism, interactions were conducted solely through spoken communication, with the chat panel from Inworld disabled. Participants communicated with the ECAs using their headset microphones and received responses through the headset speakers. Inworld provided character animation, incorporating basic states like idle and talking, along with lipsync capabilities and various gestures. These animations were activated by events tied to the dialogue context sent by the Inworld server during interactions. This system encompassed both body movements as well as facial expressions. See Supplemental video for examples.

Inworld uses personality prompts to guide the AI's language model, helping it generate responses that align with the character's unique personality and age-group. For variety in the interactions, we created six female embodied conversational agents, each appearing in a different trial (with the order of appearance randomised) and each character was generated using identical personality prompts to ensure that agent personality would not have an effect. While all the characters shared the same personality and an outfit of casual jeans and a plain t-shirt (to minimise social biases that might arise from different clothing styles \cite{oh2020economic}), they differed in hairstyle, hair color, skin tone, and voice to avoid repetition.

Each character was given the information about the name of the painting and artist that they would be standing beside (see Section \ref{sec:Procedure}), and all dialogue was generated dynamically. The age-range was selected as ``young adult," so that the speech reflected the expected age range of participants, who were university students. The personality prompts were refined over several iterations of informal pilot testing to balance friendliness with brevity of response, where the final personality prompt was: 
\begin{quote}
“Helen is very calm and reserved. Helen is next to `The Swing' by Jean-Honoré Fragonard.”
\end{quote} (The name and painting varied for each character).

\subsection{Proximity Measurement}
In each trial, participants were tasked with obtaining a specific fact from the ECA, as described in Section \ref{sec:Procedure}. Participants were instructed to initiate the conversation once at a comfortable distance from the ECA and were free to end it once they had obtained the necessary information. However, they were instructed to wait for the ECA to complete its final response before turning or moving away.

Interpersonal distance was measured as the euclidean distance from the user's HMD to the agents head, recorded every 0.5 seconds over the course of the interaction.  For measurement purposes, the course of the interaction was calculated from the beginning of the ECA’s first reply to end of their last. Participants completed three interactions, and the average interpersonal distance across these interactions was used for analysis.

While prior research has used the minimum distance~\cite{bailenson2003interpersonal,takahashi2022interpersonal}, we opted for an approach that averaged the distance over the whole interaction, which we felt was more ecologically valid and would be robust against participants accidentally moving too close to the agent (similar to recent work by Miller et al.~\cite{miller2023large}).

\section {EXPERIMENT DESIGN}

\subsection{Participants}
Participants were students undertaking a module on emerging technologies in the university. Participants undertook the experiment voluntarily without compensation. The study was approved by the Trinity College Dublin Research Ethics Committee.

70 volunteers (44 males, 24 females, 1 non-binary, 1 undisclosed) took part in this experiment.
An a priori power analysis was conducted using G$\ast$Power version 3.1.9.7 \cite{faul2007g} to determine the minimum sample size to test our primary hypothesis. Results indicated the required sample size to achieve 80\% power for detecting a medium effect, at a significance criterion of $\alpha$ = .05, was N = 48. Thus, the obtained sample size is more than adequate to test the study hypothesis.

The two participants who identified as ``prefer not to say" (n=1) and ``non-binary" (n=1) were excluded from analysis related to gender effects.
Although this decision may reduce the representation of gender diversity in the study, the small sample sizes would not have allowed for meaningful conclusions or adequately reflected the range of gender identities. These 2 participants were included in all analyses not related to gender effects.

44 participants were European, 8 were Asian, 4 Indian and 4 were African. Only 1 participant was Middle Eastern. 5 reported their ethnicity as mixed and 4 did not report their ethnicity. Participants' age ranged from 19 to 65, with a mean of 21.4, and a standard deviation of 7.73.

In terms of experience, 27 participants had above average experience with VR (24 slightly above/above average, 3 expert). 23 participants had average experience. 18 participants had below average experience (11 slightly below/below average, 7 novice.) 2 participants declined to answer on their experience with VR.

\subsection{Procedure}
\label{sec:Procedure}
Participants were initially presented with an information sheet and consent form to read, followed by a request for their signature. They also filled out a form about their demographics (age, gender, nationality, experience with VR) prior to the experiment. The participants then read some short instructions and the necessary controller inputs were shown to them. Then, they were asked to put on the HMD and hold the controllers.

Before each block, the participant started in the same real-life position and orientation marked out on the floor. Upon starting, the HMD was reset to the same virtual position and orientation so participants always had the same starting view throughout. 

Before each block, they underwent a training phase, where they were told to either walk or teleport around the room to get used to the locomotion method of that block. There was no time limit to this phase, and they could proceed whenever they wished.

The experiment involved interactions with an embodied conversational agent. The mode of interaction was active, requiring participants to initiate the verbal conversation. Our goal was to create a natural interaction in an art gallery setting, where participants could engage with a helpful gallery assistant to learn more about a painting they had just viewed. The participant was shown a screen with a painting and a fact to find out about the artwork e.g. \textit{``What time period is the painting from?''} Once they were finished viewing the screen they clicked a button and were told to do the following: approach the character until they were at a comfortable distance from them, then proceed to have a conversation with the agent - in which they were to find out the fact about the painting. The agent originally faced the center of the room and would turn to face the user if they entered its 120-degree field of view. The painting and fact were not visible after the participant clicked ready, to prevent the participant turning to look at the painting mid conversation and interfere with proximity measurements. The participant was told they could end the conversation when they wished; upon doing so they should return to press a button to progress onwards. Participants were told not to move away before the interaction had finished, to prevent interference to proximity measurements. The experiment involved two blocks in randomised order, walking and teleportation. Each trial had 3 repetitions, in which the character, painting and fact changed each time for variety. The position of the character also alternated from each end of the virtual room to encourage greater use of locomotion.

After each block, the participants filled out a questionnaire about: their feelings of togetherness with the embodied agent (co-presence), control of the virtual body (agency) and feeling of ownership of the virtual body (Body Ownership). 
The questions relating to Body Ownership \textit{(BO)} and Agency \textit{(AG)} were based on \cite{podkosova2018co}'s adaption of \cite{gonzalez2018avatar}, with adjustments to agency questions made for time constraints. Similarly, to assess co-presence \textit{(CP)} we used \cite{podkosova2018co}'s  adaption of \cite{biocca2003toward}.

The questionnaire also included two additional questions: a question about their anxiety about colliding \textit{(COL)} with the agent (from \cite{podkosova2018co}). We also felt the participants' self-reported accuracy \textit{ (ACC)} would be of interest, so we also questioned participants on how easy they felt it was to get into place to converse with the character. All questions are included in Table \ref{table:questions}. All questions are answered on a 7-point Likert scale from \textit{strongly disagree} to \textit{strongly agree}.

\begin{table*}[h]
  \centering 
  \caption{Questionnaire including the following dimensions: Co-Presence (CP), Agency (AG), Body Ownership (BO), Accuracy (ACC) and Collision (COL). Regarding questions about the virtual body, participants were informed that these referred to the body they observed when looking down, to avoid confusion with the agent's body.}
  \label{table:questions}
  \begin{tabular}{|l|p{0.9\linewidth}|}
    \hline
    \textbf{CP1} & To what extent did you have the feeling of the AI character being together with you in the virtual environment? \\
    \hline 
    \textbf{CP2} & The sense of being together with the AI character resembles the sense of being with others in the real world. \\
    \hline  
    \textbf{AG1} & It felt like I could control the virtual body like it was my own. \\
    \hline  
    \textbf{AG2} & I felt as if the virtual body was moving by itself. \\
    \hline  
    \textbf{AG3} & I felt as if the movements of the virtual body were influencing my own movements. \\
    \hline  
    \textbf{BO1} & I felt as if the virtual body was my own body. \\
    \hline  
    \textbf{BO2} & It felt as if the virtual body was someone else. \\
    \hline  
    \textbf{BO3} & It seemed as if I might have more than one body. \\
    \hline  
    \textbf{COL} & To what extent were you worried that you would collide with the character? \\
    \hline  
    \textbf{ACC} & I found I could easily get in place to converse with the character. \\

    \hline

  \end{tabular}
\end{table*}

After the final block, the participants were asked additional questions: they were also asked which locomotion method they preferred and found easier to use. They were asked for additional comments on the AI characters, walking and teleportation in an open question box, which could be left blank. 

Each participant performed all trials. They were given the opportunity to take breaks in-between each trial. They were also told they could stop and withdraw from the experiment at any time.

\section{Results}
To analyse the results of proximity, we conducted a repeated-measures ANOVA with a within-subject factor of locomotion method (walking or teleportation). We also included between-subjects factors of participant gender and trial order in the analysis.
The data is assumed to be normal under the Shaprio Wilk and Lilliefors test but not under 
the Kolgogorov Smirnov test. As such we cannot conclude the data to be normal. However the ANOVA is robust to non-normality in larger samples (n=70).
Based on visual inspection through a boxplot and also by checking for values outside the range [Q1 - 3 × IQR, Q3 + 3 × IQR], no extreme outliers were observed in the data. As the repeated measure had two levels (teleportation and walking) the sphericity assumption was automatically met. The data was homoscedastic between each level of locomotion (Levene test). The average interpersonal distance of all trials within each block was used, thereby satisfying the independence assumption. Statistical significance is reported at $p < 0.05$.

\begin{figure}[tb]
 \centering 
 \includegraphics[width=\columnwidth]{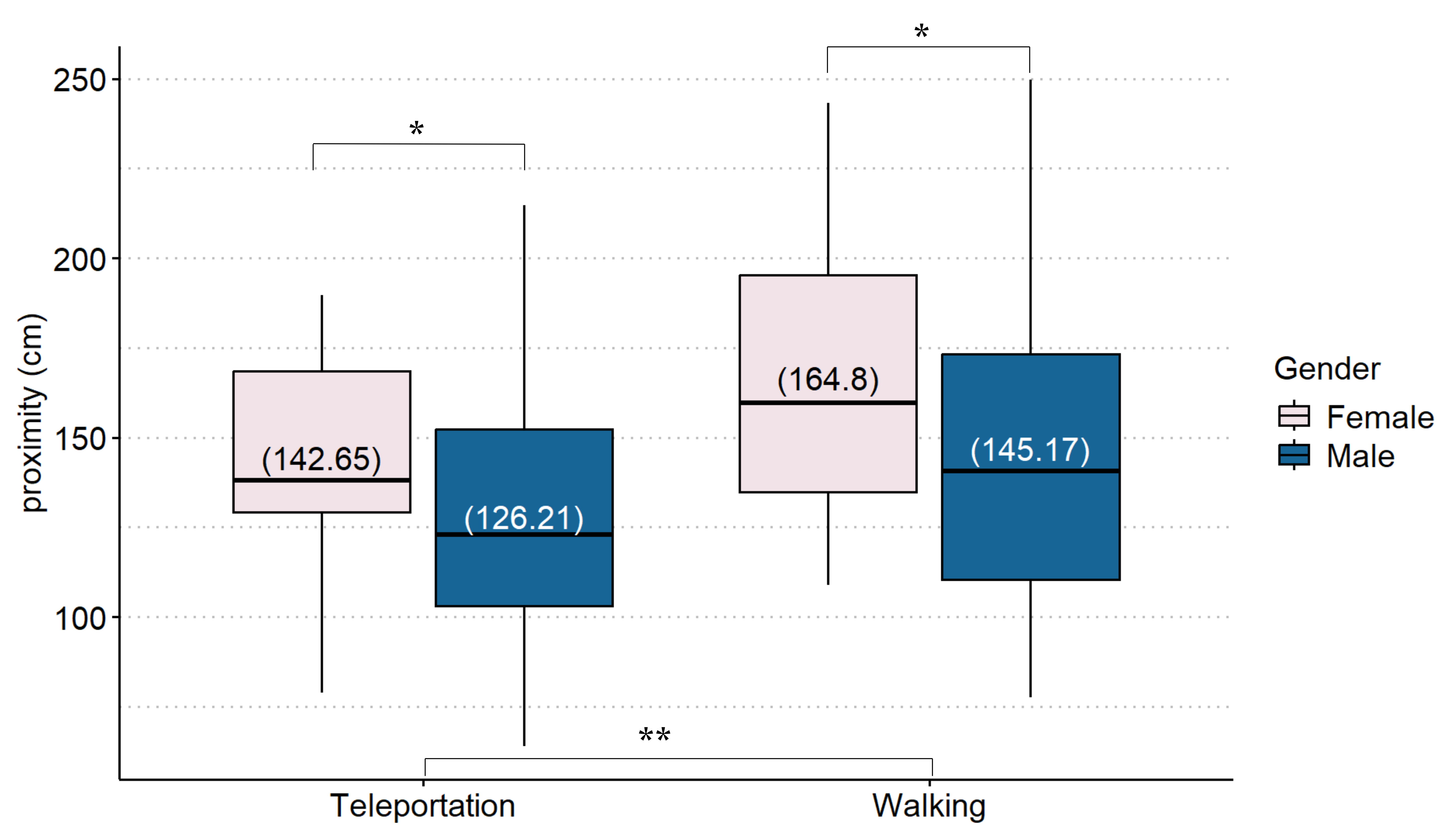}
 \caption{Boxplots of proximity measurement distribution across locomotion type and gender (mean is labelled)}
 \label{fig:boxplots}
\end{figure}

We found a significant within-subjects effect for locomotion type (F(1, 64) = 15.28, \(p < .01\), \(\eta^{2}_{p} = 0.19\), \(1 - \beta = 0.97\)). When participants used walking as the locomotion method, they interacted with the embodied agent  at a greater distance ($M : 152cm, SD : 4.06cm$) than when using teleportation ($M : 131 cm, SD: 3.37 cm$), see \cref{fig:boxplots}.

For between subject factors we found no order effects but the proximity was significantly different according to the gender of the participant  
(F(1, 64) = 5.899, \(p = 0.036\), \(\eta^{2}_{p} = 0.067\),\(1-\beta=0.56\)).
Female participants interacted with the embodied agent at a greater distance $(M : 154cm, SD : 3.38cm)$ than male participants $(M : 136cm, SD : 3.48cm)$, see \cref{fig:boxplots}. There were no interaction effects.

\subsection{Questionnaire responses}

For agency, questions \textit{AG1}, \textit{AG2} and \textit{AG\_AVG} were significantly higher for walking than for teleporting. \textit{BO1}, \textit{BO3}, and \textit{BO\_AVG} were also significantly higher for walking than teleportation. Co-presence did not significantly differ for each locomotion type. \textit{ACC} was significantly higher for walking than for teleportation. Finally, collision anxiety (\textit{COL}) was significantly higher for teleportation compared to walking.  All test statistics are included in \cref{table:q_results}.

\newcolumntype{L}{>{\raggedright\arraybackslash}X}  
\newcolumntype{C}{>{\centering\arraybackslash}X}   
\newcolumntype{R}{>{\raggedleft\arraybackslash}X}  

  \begin{table*}[h]
  \caption{Results of Wilcoxon Signed Rank Tests with mean for each locomotion group, sum of positive and negative ranks, Z, P, FDR Adjusted p-values and effect size r. }
  \label{table:q_results}
  \scriptsize%
  \centering%
 \begin{tabularx}{0.83\textwidth}{ 
   R
    p{0.7cm} 
    p{0.7cm} 
    p{1.5cm} 
    p{1.5cm} 
    p{0.8cm} 
    p{1.2cm} 
    p{2cm} 
    p{1cm} 
}

   \textbf{Question} & \textbf{mdW} & \textbf{mdT} & \textbf{Sum+ Ranks} & \textbf{Sum- Ranks} & \textbf{z} & \textbf{P} & \textbf{FDR Adjusted P} & \textbf{r} \\
    \hline
    \textbf{*AG1} & 6.0 & 4.9 & 72.5 & 1152.5 & 5.42 & $<0.01^*$ & $*<0.0433$ & 0.65 \\
    \textbf{*AG2} & 5.6 & 4.6 & 176.5 & 858.5 & 3.87 & $<0.01^*$ & $*<0.0325$ & 0.46 \\
    AG3 & 4.1 & 4.3 & 626.5 & 501.5 & 0.68 & 0.499 & 0.499 & 0.08 \\
    \textbf{*AG\_AVG} & 5.2 & 4.6 & 385.5 & 1630.5 & 4.73 & $<0.01^*$ & $*<0.026$ & 0.57 \\
    CP1 & 5.2 & 5.0 & 272.5 & 468.5 & 1.47 & 0.142 & 0.1676 & 0.18 \\
   CP2 & 4.0 & 3.7 & 239.5 & 463.5 & 1.73 & 0.083 & 0.1079 & 0.21 \\
    CP\_AVG & 4.6 & 4.4 & 378.5 & 749.5 & 1.98 & 0.047* & 0.0678 & 0.24 \\
    \textbf{* BO1} & 4.9 & 4.4 & 358.0 & 867.0 & 2.59 & 0.010* & $*0.0186$ & 0.31 \\
    BO2 & 4.7 & 4.5 & 292.0 & 488.0 & 1.40 & 0.161 & 0.1744 & 0.17 \\
    \textbf{* BO3} & 5.5 & 5.2 & 288.0 & 615.0 & 2.10 & 0.036* & $*0.0585$ & 0.25 \\
    \textbf{* BO\_AVG} & 5.0 & 4.7 & 541.0 & 1289.0 & 2.78 & 0.005* & $*0.0108$ & 0.33 \\
    \textbf{* ACC} & 6.3 & 5.3 & 187.5 & 1037.5 & 4.30 & $<0.001^*$ & $*<0.013$ & 0.51 \\
    \textbf{* COLL} & 2.7 & 4.1 & 1029.5 & 146.5 & 4.56 & $<0.001^*$ & $*<0.0065$ & 0.54 \\
  \end{tabularx}%
\end{table*}

\subsubsection{Gender Differences \& Participant Preferences}
For both locomotion methods we compared the same questionnaire data across participant gender using the Mann Whitney U Test (as the sample sizes were unequal). Using the false discovery rate for correction, no results were significant.

We also asked which locomotion method participants preferred and found easier to use. Most ($n=58, 83\% $ participants) preferred walking as a locomotion method and also found it easier to use ($n=56, 80\%$).

\subsection{Subjective Comments}
Participants were invited to give additional comments on teleportation, walking, and their experience with the agent. There were 66, 64, and 62 responses respectively (comments were optional). These were reviewed informally using approaches inspired by Braun \& Clarke \cite{braun2012thematic}.
From this analysis several themes emerged such as distance, speed or immersion. In many cases themes were further divided into subthemes. We found that teleportation prompted many reports of distance estimation struggles, while users generally reported distance estimation benefits for walking. For speed, they generally had positive comments about teleportation as it was fast, easy and convenient, while walking was generally considered slower but also easy. For immersion, users found walking to be more like real life, where they were less aware of VR, while they found teleportation to be more like a video-game and less natural. The full details of this are included in Supplemental Material Tables 1--3.

As the locomotion was the primary focus of our study we classified the ECA comments simply by general sentiment, as ``Positive" (53\%, n=33), ``Negative" (19\%, n=12) and ``Neutral/Both" (0.27\%, n=17). Full details in Table 4 of Supplemental Material.

\section{Discussion}
In this study, we examined the impact of teleportation on IPD in virtual reality, focusing on interactions with embodied conversational agents. Our main finding is that IPD differed when participants approached an agent using teleportation compared to natural walking, thus supporting our primary hypothesis (\textbf{H1}). More specifically, users got closer to the agent after teleportation than walking. We propose that this difference may be attributed to distance estimation biases and increased spatial cognitive load introduced by teleportation.

\subsection{Perspective and Measurement Planes}

Spatial perception biases in VR influence distance estimation based on perspective \cite{lin2014distance, renner2013perception} and the plane of measurement \cite{geuss2012effect}. In our experiment, we hypothesise that users initially choose a teleportation destination by estimating their preferred IPD from the agent. This estimation is based on an exocentric (object-to-object) perspective for evaluating distance. Upon arriving near the agent, users reassess the distance from an egocentric (self-to-object) perspective. In contrast, walking allows for continuous egocentric evaluation, enabling users to refine their distance estimations progressively as they approach the agent.

Exocentric distances in VR have been reported to be overestimated \cite{waller1999factors,wartenberg2003precision}, suggesting that users might position themselves too close to the agent during teleportation, resulting in a reduced IPD. Conversely, egocentric distances are typically underestimated \cite{lin2014distance}, indicating that walking might lead to a larger IPD. This could explain the variation in IPD between locomotion methods. Although some studies report conflicting results regarding exocentric distance estimation in VR \cite{geuss2012effect, li2011underestimation, lin2014distance}, the consistent underestimation of egocentric distances supports the likelihood of a discrepancy in IPD perception across locomotion methods.

Furthermore, measurement planes can introduce biases in distance estimation and may influence IPD. When walking towards the ECA, distance estimation is generally made from an egocentric perspective in the depth plane. With teleportation, perceived distance can vary depending on the alignment of the teleportation destination with either the depth or width plane, potentially influenced by the ECA's orientation. A study by Guess \cite{geuss2012effect} found that distance measurements in VR are accurate in the width plane but are often underestimated in the depth plane. Their results suggest that distance underestimation in VR occurs primarily in the depth plane. Since walking predominantly involves depth-plane assessments and teleportation involves some width-plane measurements, users might experience a larger IPD during walking compared to teleportation. Our results demonstrated that teleportation reduces IPD compared to walking, suggesting that the measurement plane in each method may contribute to distance estimation biases. An important consideration is that a user's teleport destination can be oriented in various ways depending on how they position themselves. Research indicates that distance estimations become less accurate for objects further from the center of the field of view \cite{peillard2019virtual}, meaning the positioning of the teleport destination adds complexity to distance estimation, warranting further research. 

Previous studies suggest that the absence of vestibular and proprioceptive feedback during teleportation may lead to underestimation of distances \cite{campos2012multisensory}. Based on this, one might expect teleportation to result in an increased IPD. However, our findings did not show an increase in IPD. Teleportation often requires estimating distances in extrapersonal space, where underestimation is common \cite{armbruster2008depth, naceri2011depth, ng2016depth}, while peripersonal space tends to be more accurately estimated, though occasionally overestimated \cite{armbruster2008depth, naceri2011depth}. Despite this, our findings did not show an increase in IPD.

A study by Keil et al. \cite{keil2021effects} proposed that both the reliance on extrapersonal distance and the absence of self-motion cues in teleportation would lead to distance underestimation. While Keil's study confirmed that distances are underestimated during teleportation, it found no significant differences in estimation compared to continuous locomotion. Importantly, Keil’s research focused on estimating the distance travelled, not IPD, and used artificial locomotion rather than natural walking. In contrast, our study involved natural walking and measured IPD, not distance travelled. Since walking also requires some degree of extrapersonal distance estimation, any proximity biases related to different locomotion methods might offset each other. Our results suggest that factors such as perspective shifts and measurement planes have a more significant impact on IPD than proximity biases or self-motion cues.

In summary, the observed reduction in IPD during teleportation may be shaped by biases related to spatial processing differences during the initial teleport. The frame of reference (perspective) and the alignment of the teleportation destination (measurement plane) could influence IPD.

\subsection{Spatial Cognitive Load}
\label{sec:cognitiveload}

As discussed, we believe that differences in distance estimation biases impact IPD because the initial assessment of distance before teleporting can be influenced by distance estimation biases. However, once users teleport close to an agent, they ultimately obtain a viewpoint comparable to walking up to the agent. Given this, one would expect that users would recognise any inaccuracies in their initial distance estimation. Regardless of how they judged the distance prior to teleporting, it would be reasonable to expect them to adjust their position accordingly, leading to similar IPDs across both locomotion methods. However, this did not occur in our study. Users may have felt uncomfortable but chose to interact at an unnatural distance due to spatial cognitive costs of teleporting \cite{cherep2020spatial} or the high task complexity of using a controller to adjust \cite{laviola2001hands}. This cognitive load might lead users to accept the uncomfortable closeness rather than exerting the additional effort to adjust their position. Although we did not formally analyse the locomotion strategies used by participants, we observed general patterns in their behavior. Participants typically made one teleport and only re-teleported if they were at extreme distances from the agent (either too close or too far). If the agent was visible and the distance seemed acceptable, they accepted their position. User comments indicate a tendency to overshoot and discomfort with being `too close' to the agent, implying that their preference for interpersonal distance did not differ across each locomotion method. Despite having the option to adjust (there was no limit on number of teleports), users often interacted at a closer proximity.

If a user teleports too far from the agent, they might try to correct by teleporting forward, risking further overshooting with each subsequent teleport. Correcting overshooting may involve a small backward step, which may feel unsafe due to potential collisions with real-world objects. Alternatively, teleporting backward requires significant effort, either by twisting around or blindly choosing a rearward destination. As a controller based action, task complexity is shifted from the legs to the hands which has been argued to increase cognitive load \cite{laviola2001hands}. Additionally, teleportation incurs spatial cognitive costs \cite{cherep2020spatial}. Thus, the effort for minor adjustments may not be worthwhile. 

In summary, we propose that the observed IPD effects arise from an initial distance misjudgment during teleportation and a subsequent reluctance to adjust. Due to the complex nature of distance estimation in VR versus real-world scenarios \cite{lampton1995distance}, pinpointing a single cause for this misjudgment is challenging. Further research is needed to clarify the interactions among perspective, measurement plane, self-motion cues, and proximity biases across different locomotion methods.

\subsection{Findings from Subjective Comments}
Although open-ended questions may be criticised for subjectivity, thematic analysis provides a flexible  way to explore the themes revealed in participant responses \cite{braun2012thematic}. In this study we found user comments to be extremely insightful. Notably, many users mentioned distance estimation in locomotion comments even though they were asked an open-ended question without specific prompts. In the walking condition, comments mentioned distance estimation benefits, such as \textit{``it was easy to feel the distance"} and \textit{``you can control the distance at which you stand from the character"}.

In contrast, teleportation prompted many reports of distance estimation struggles \textit{``I sometimes overshot where I wanted to teleport"} and \textit{``it was hard to judge distance as accurately as walking"}.

We noticed that comments about discomfort and feeling ``too close" were prevalent in teleportation scenarios but not in walking scenarios. This suggests that the IPD maintained during walking felt natural to users, while it felt unnatural during teleportation. The frequency of distance estimation in user comments further supports our argument that distance estimation biases contributed to the observed effect on IPD.

\subsection{Embodied Conversational Agents}
To our knowledge, this experiment is the first to explore proximity in VR during interactions with ECAs. Our study introduces a promising approach to measure proximity in a more naturalistic context. Traditional proximity research often involves participants approaching or being approached by a silent virtual human, which makes the study’s intentions clear and may influence results. We believe that incorporating ECAs into proximity studies holds significant potential. Additionally, the use of prompt generation to create ECAs enables the manipulation of variables such as personality, allowing for the assessment of their effects on personal space.

We would expect our findings to remain consistent with a uniform interactant. However, to further explore the potential of ECAs for proximity studies, it would be valuable to investigate how personal space management might vary in user-user versus user-agent interactions.

The artificial nature of the agents used in this study, along with the potential influence of the Uncanny Valley effect, requires careful consideration. Subtle aspects such as the agents' motion or voice quality may have influenced participants' spatial behaviours in ways that are not immediately apparent.

Although our agents were distinctly AI-generated, they maintained consistent designs across both teleportation and walking conditions. This consistency suggests that proximity differences are unlikely to be attributed to variations in the agents themselves. However, the nuanced role of the Uncanny Valley effect in affecting spatial behaviors is not fully understood, remaining open for further exploration.


\subsection{Other Findings} 
The results revealed some other interesting findings. A stronger sense of agency and body ownership was reported by participants when walking compared to teleportation, supporting \textbf{H2} and aligning with previous work \cite{leonardis2014multisensory}. We found no difference in co-presence with the agent between natural walking and teleportation, contrary to \textbf{H3}. We had expected higher co-presence in the walking condition based on Freiwald’s findings that continuous joystick movement yields greater co-presence than teleportation \cite{freiwald2021effects}. However, our study is the first to compare natural walking and teleportation specifically, rather than joystick-based locomotion. Future work could explore the specific mechanisms behind each locomotion method to better understand how they individually contribute to feelings of co-presence. Additionally, co-presence could be more reliant on the interaction itself rather than locomotion type. As expected (\textbf{H4}), female participants kept a larger distance from the agents than males, a gender effect that is observed in previous studies \cite{iachini2016peripersonal, zibrek2020effect}.

The average proximity for both locomotion methods fell in the close social zone, as defined by Hall \cite{hall1966hidden}. The agent-participant relationship would be expected to fall here as this zone would be common for new acquaintances or within casual social gatherings. Future work could investigate if the observed effect of locomotion holds among participants with pre-existing relationships i.e., interactions that take place in the personal zone.

Walking was overwhelmingly preferred and was also reported as easier to use. As expected, participants reported feeling a higher sense of accuracy when walking compared to teleportation. Participants expressed a higher level of concern about colliding with the agent while teleporting compared to walking. These findings were also reflected in the comments the participants made about each locomotion method.

There were no gender differences in any questionnaire items for either locomotion method. While some studies have shown mixed gender effects on co-presence—such as females reporting lower levels compared to males \cite{freiwald2021effects}, or the opposite \cite{podkosova2018co}—these studies involved co-presence with another user, not an agent. Our study found no gender differences in co-presence with the ECA, though the novelty of the ECA might have obscured any potential differences. Additionally, no significant gender differences were found for agency or body ownership, aligning with a review by Mottelson \cite{mottelson2023systematic}.

\section{Limitations and Future Work}
Although we have provided explanations for the effect observed on IPD, our results are limited to the context of our study.

For example, we cannot rule out that the specific implementation of teleportation used was responsible for the effect observed on IPD. Testing this effect with various implementations of destination recticles, trajectory lines, or transition types during teleportation would be an ideal extension to the study. Future studies should also explore whether the observed effects are exclusive to the comparison between natural walking and teleportation, or if they extend to other continuous versus non-continuous locomotion methods.

We used the MetaMovement SDK to infer participants' body poses instead of employing motion trackers. While this method offers convenience and ease of setup, it may have influenced the accuracy of body pose data, potentially impacting the results. However, we believe this did not have an effect, as we did not receive many participant comments regarding issues with body tracking. Since body ownership depends on the artifical body and its movement \cite{kilteni2012sense}, the use of this SDK may have influenced this aspect of our study. Yet, as expected, we did find a significant difference in body ownership between methods. Still, future research should validate the MetaMovement SDK’s accuracy against dedicated motion tracking systems to ensure robust findings.

Additionally, a potential limitation of our study is the controlled experimental setting. Although participants were not given a time limit, they may have felt pressured by the environment, which could have influenced their ability or effort to adjust distance estimates as they might in a more natural setting.

Future advancements in LLMs may introduce greater variability to ECAs; however, we believe our work is currently reproducible due to the generic nature of the agents' responses. Nonetheless, evolving AI technologies with better conversational abilities and more natural voice and movement capabilities could make replication more difficult in the future. The artificial nature of the agents used in our study could also have affected co-presence. Therefore, repeating this study without conversation or with avatars would be beneficial. Future research could explore other agent types, like non-conversational virtual agents or confederates, to examine different  interaction types. 

All of our ECAs were female and generated using identical prompts, sculpted to make the ECAs friendly and welcoming to users, using the same cartoonish style for each one. As such, our study did not examine effects such as stylisation and personality - which makes for interesting avenues for further investigation. It also remains unclear how participants would respond to male agents. Studies suggest that female agents are often perceived as friendlier than male agents \cite{abele2003dynamics}, so exploring the combination of AI novelty with male agents could yield intriguing insights. As proximity is an important non-verbal communication in VR, future studies could explore further the use of ECAs to facilitate social interactions.

In our experiment, participants were instructed to approach the agent until they were at a comfortable distance from them before starting the conversation. The purpose of this was to prevent interaction while moving through the virtual environment. This phrasing, `approach until comfortable' or `be approached until uncomfortable', is typical in VR proximity studies (including \cite{iachini2014body, zibrek2020effect, iachini2015influence}) but may have hinted at the study's purpose and influenced behavior. Nevertheless, we do not believe this significantly affected the results. Informal feedback suggested participants suspected the experiment aimed to compare locomotion methods or interaction quality with the ECA.

Finally, investigating how people adjust when teleporting to a target destination would offer valuable insights. A formal analysis of teleportation strategies, such as the number and magnitude of teleports would help better understand participants' behavior. It would be important to understand the types of corrections individuals make, whether they involve teleporting or stepping in real life, as well as the direction and extent of these adjustments. Additionally, examining eye gaze could reveal if users tend to look at the ground when making small adjustments near their feet or focus on the agent to assess their distance from them.

\section{Conclusion}
The moderation of interpersonal distance holds importance in ensuring user comfort, especially in social VR where interactions are common. Our study compared teleportation and natural walking in VR and found that participants maintained closer distances during teleportation. Additionally, female participants kept more distance than males. Natural walking was linked to greater agency and body ownership, while co-presence remained the same. We propose that biases in distance estimation, along with a reluctance to make adjustments, contributed to the reduced distances observed during teleportation. This highlights the importance of considering locomotion methods in VR and underscores the need for more research on how these methods affect spatial perception and social interactions.

\acknowledgments{%
	This work was conducted with the financial support of the Research Ireland Centre for Research Training in Digitally-Enhanced Reality (d-real) under Grant No. 18/CRT/6224 and RADICal (Grant No. 19/FFP/6409). This project has also received funding from the European Union’s Horizon 2020 Research and Innovation Programme under the HUMAN + COFUND Marie Skłodowska-Curie grant agreement No. 945447%
}

\bibliographystyle{abbrv-doi-hyperref}

\bibliography{template}





\graphicspath{{figures/}{pictures/}{images/}{./}} 





\clearpage 
\section*{Supplementary Material}







Participants were invited to give additional comments on teleportation, walking, and their experience with the agent. There were 66, 64, and 62 responses respectively (comments were optional). Thematic analysis was conducted using approaches outlined by Braun \& Clarke \cite{braun2012thematic}. 

\subsection*{Coding}

From thematic analysis of locomotion comments (teleportation and natural walking) several themes emerged such as distance and immersion. Each of these themes was broken down into various subthemes. The themes, sub-themes, and descriptions used to code each comment are shown in Table \ref{tab:themes_subthemes}. The results 
for teleportation comments are included in table \ref{tab:teleportTA}. Results for walking comments is included in table \ref{tab:walkingTA_new}

\subsection*{Sentiment Analysis}
The primary focus of our study was locomotion. However as the ECA's were a novel aspect of our experiment we still wished to gauge participant's feelings towards them. Therefore we classified ECA comments by general sentiment, as ``Positive" (53\%, n=33), ``Negative" (19\%, n=12) and ``Neutral/Both" (0.27\%, n=17). Results are included in table \ref{table:ecas_sentiment}.

\begin{table*}[t]
  \centering
  \caption{Thematic analysis overview of themes, sub-themes and decsription }
  \label{tab:themes_subthemes}
  \begin{tabular}{|p{0.15\linewidth}|p{0.2\linewidth}|p{0.5\linewidth}|}
    \hline
    \textbf{Theme} & \textbf{Sub-theme} & \textbf{Description} \\
    \hline

    \textbf{Distance, Precision \& Accuracy} & \multicolumn{2}{|p{0.6\linewidth}|}{ \textbf{Mentions some aspect of precision, accuracy or distance related to teleportation method}} \\
    \cline{2-3}
     & Overall Negative & Mentions difficulties with estimating distance or expresses negative experiences related to the precision or accuracy of the method. This includes any impairments or challenges associated with these aspects. \\
    \cline{2-3}
    & Overall Positive & Describes ease with estimating distance or shares positive experiences related to improved precision or accuracy of the method. \\
    \cline{2-3}
    & Overall Mixed & Expresses a combination of both positive and negative sentiments regarding distance estimation, precision, or accuracy related to the teleportation method. \\
    \cline{2-3}
    & Distance Estimation Issues & Specifically mentions difficulties in estimating or gauging distance, such as uncertainty about how far they would travel. \\
    \cline{2-3}
    & Distance Estimation Advantages & Describes ease or advantages related to estimating or gauging distance, including a clear sense of how far one would travel. \\
    \cline{2-3}
    & Overshooting or Too Close & Describes experiences of teleporting too close to a character or object, or tendencies to overshoot and collide. If they also mention going too far, apply the corresponding code. \\
    \cline{2-3}
    & Too Close or Too Far & Expresses a tendency to both overshoot or get too close, as well as undershoot or end up too far from the intended target. \\
    \hline

       \textbf{Speed, Ease \& Convenience} & \multicolumn{2}{|p{0.6\linewidth}|}{ \textbf{Mentions some aspect of the speed, ease or convenience of the method}} \\
          \cline{2-3}
          
    & Fast & Describes the speed of the method as fast or equivalent. \\
    \cline{2-3}
    & Slow & Describes the speed of the method as slow or equivalent. \\
    \cline{2-3}
    & Easy or Convenient & Highlights the ease or convenience of the teleportation method, such as the simplicity of controls or the overall user experience. \\
    \hline

        \textbf{Immersion} & \multicolumn{2}{|p{0.6\linewidth}|}{ \textbf{Mentions aspects related to immersion, including the sense of realness, naturalness, or gamelike qualities of the teleportation method.}} \\
          \cline{2-3}
     & Negative Impact & Describes negative effects on immersion, including reduced naturalness or realness. This also includes descriptions of the method as ``gamelike'' or having a game-like quality. \\
    \cline{2-3}
    & Mixed Impact & Provides a combination of positive and negative views on immersion, realness, naturalness, or gamelike qualities. For example, if a comment indicates that the method felt both more natural and game-like, classify it as mixed. \\
    \hline

         \textbf{Fun} & \multicolumn{2}{|p{0.6\linewidth}|}{ \textbf{Mentions aspects of enjoyment or interest related to the locomotion method.}} \\
          \cline{2-3}
     & Fun/Interesting & Expresses that the locomotion method was fun, enjoyable, cool, or interesting, or that it was more enjoyable compared to alternative locomotion methods. \\
    \cline{2-3}
    & Less or Not Fun & Expresses that the locomotion method was not fun, enjoyable, cool, or interesting, or that it was less enjoyable compared to alternative locomotion methods. \\
    \hline

         \textbf{Other} & \multicolumn{2}{|p{0.6\linewidth}|}{ \textbf{Any other sub-theme}} \\
          \cline{2-3}
     & Learning Curve & Refers to improvements in skill or proficiency with the teleportation method as users gain more experience over time. This includes becoming better or more adept with repeated use. \\
    \cline{2-3}
    & Motion Sickness & Any mention of experiencing sickness, discomfort, or nausea related to the use of the teleportation method. \\
    \cline{2-3}
    & Practical for Space Constraints & Mentions how the teleportation method is suited for situations with limited physical space or constraints, emphasizing its practicality in such environments. \\
    \cline{2-3}
    & Smooth/Seamless/Fluid & Refers to how smooth, seamless, or fluid the method feels during use, including aspects like continuous motion and lack of interruptions or jerks. \\
    \cline{2-3}
    & Collision Anxiety & Any mention of anxiety or concern about potential collisions with real-life objects or characters while using the method. \\

    \hline
  \end{tabular}
\end{table*}

\begin{table*}[t]
  \centering 
 \caption{
  Thematic analysis overview of 66 comments on \textit{Teleportation}: with \textit{theme}, \textit{sub-themes}, number of comments (\textit{n}), and the proportion of supplied comments (\textit{p}) to quantify feedback. Note that overlapping themes may mean that the proportions do not sum up to a total, as some comments might belong to multiple themes and sub-themes.
}
  \label{tab:teleportTA}
  \begin{tabular}{|p{0.15\linewidth}|p{0.15\linewidth}|p{0.58\linewidth}|p{0.08\linewidth}|}  
    \hline
    \textbf{Theme \textit{(n, p)}} & \textbf{Subtheme} & \textbf{Examples} & \textbf{\textit{n, p}} \\ 
    \hline

    \multirow{5}{*}{\parbox{1\linewidth}{\RaggedRight\textbf{Distance, Precision \& Accuracy }(22,~0.33)}}  
    & Overall Negative & ``did not offer as much precision on where I was going to land" & 21, 0.32 \\ 
    \cline{2-4}
    & Overall Mixed & ``It is very precise but I had to calibrate myself before I felt comfortable moving around" & 1, 0.02 \\ 
    \cline{2-4}
    & Distance Estimation Issues & ``I judged the distance wrong', ``the distance gauging was off", ``hard to judge distance" & 8, 0.12 \\ 
    \cline{2-4}
    & Overshooting or Too Close & ``Brought me on top of the characters", ``teleporting too close", ``further than intended" & 10, 0.18 \\ 
    \cline{2-4}
    & Too Close or Too Far & ``I went too far or too short" & 2, 0.03 \\ 
    \hline
  
    \multirow{3}{*}{\parbox{0.8\linewidth}{\RaggedRight\textbf{Speed, Ease \& Convenience } (24, 0.36)}}  
    & Fast & ``quick", ``fast", ``instantly zip around the room" & 12, 0.18 \\ 
    \cline{2-4}
    & Slow & ``took more time", ``hard to move quickly" & 2, 0.03 \\ 
    \cline{2-4}
    & Easy or Convenient & ``easy", ``convenient" & 17, 0.26 \\ 
    \hline
    
    \multirow{2}{*}{\parbox{0.15\linewidth}{\RaggedRight\textbf{Immersion }(15,~0.23)}}  
    & Negative Impact & ``less natural", ``felt more like a video game" & 14, 0.21 \\ 
    \cline{2-4}
    & Mixed Impact & ``Felt more natural than walking as it felt like a game-like environment." & 1, 0.02 \\ 
    \hline

    \multirow{3}{*}{\parbox{0.15\linewidth}{\RaggedRight\textbf{Fun}~(7,~0.11)}}  
    & Fun/Interesting & ``fun", ``interesting" & 6, 0.09 \\ 
    \cline{2-4}
    & Less or not fun & ``less fun" & 1, 0.02\\ 
    \cline{2-4}
    \hline

    \multirow{4}{*}{\parbox{0.15\linewidth}{\RaggedRight\textbf{Other} (11,~0.17)}}  
    & Learning Curve & ``After some tries it's easy to get used to", ``got the hang of it" & 4, 0.06 \\ 
    \cline{2-4}
    & Motion Sickness & ``sickening", ``motion sickness" & 2, 0.03\\ 
    \cline{2-4}
    & Obstacle Avoidance & ``I knew I wouldn't walk into something", ``not so scary that you won't bump into something." & 5, .08 \\ 
     \cline{2-4}
   & Control & ``it is slightly harder to be in control in comparison to walking", `` I felt like I had less control." & 3, .08 \\ 
    \hline

  \end{tabular}
\end{table*}

\begin{table*}[t]
  \centering 
 \caption{
  Thematic analysis overview of 64 comments on \textit{Walking}: with \textit{theme}, \textit{sub-themes}, number of comments (\textit{n}), and the proportion of supplied comments (\textit{p}) to quantify feedback. Note that overlapping themes may mean that the proportions do not sum up to a total, as some comments might belong to multiple themes and sub-themes.
}
  \label{tab:walkingTA_new}
  \begin{tabular}{|p{0.15\linewidth}|p{0.15\linewidth}|p{0.58\linewidth}|p{0.08\linewidth}|}   
    \hline
       \textbf{Theme \textit{(n, p)}} & \textbf{Subtheme} & \textbf{Examples} & \textbf{\textit{n, p}} \\ 
    \hline
 
    \multirow{4}{*}{\parbox{1\linewidth}{\RaggedRight\textbf{Distance, Precision \& Accuracy} (10,~0.16)}}  
    & Overall Positive & ``very accurate", ``able to be more precise" & 9, 0.14 \\ 
    \cline{2-4}
    & Overall Negative & ``hard to tell how far you were walking" & 1, 0.02 \\ 
    \cline{2-4}
    & Distance Estimation Issues & ``hard to tell how far you were walking" & 1, 0.02 \\ 
    \cline{2-4}
    & Distance Estimation Advantages & ``easy to feel the distance", ``easy to find a comfortable distance", ``control the distance" & 5, 0.08 \\ 
    \hline

    \multirow{2}{*}{\parbox{0.8\linewidth}{\RaggedRight\textbf{Speed \& Ease}~(12,~0.19)}}  
    & Easy & ``easy" & 11, 0.17 \\ 
    \cline{2-4}
    & Slow & ``takes time to move around" & 1, 0.02 \\ 
    \hline

       \multirow{2}{*}{\parbox{1\linewidth}{\RaggedRight\textbf{Immersion} (25,~0.39)}}  
    & Positive Impact & ``less aware of VR", ``like real life" & 24, 0.38 \\ 
    \cline{2-4}
    & Negative Impact & ``I was extremely wary of my position in the real room and thought about that more than where I was in the virtual room" & 1, 0.02 \\ 
    \hline

    \multirow{2}{*}{\parbox{0.8\linewidth}{\RaggedRight\textbf{Other} (16,~0.25)}}  
    & Smooth/Seamless/Fluid & ``very fluid", ``seamless" & 5, 0.08 \\ 
    \cline{2-4}
    & Collision Anxiety & ``conscious of the available space", ``nervous about boundaries in the real room" & 11, 0.17 \\ 
    \hline
  \end{tabular}
\end{table*}

\begin{table*}[t]
  \centering 
\caption{Sentiment Analysis of Conversational Agents (ECAs) with Sentiment, Description, Examples, number of comments (\textit{n}), and the proportion of supplied comments (\textit{p}) to quantify feedback}
\label{table:ecas_sentiment}
  \begin{tabular}{|p{0.095\linewidth}|p{0.38\linewidth}|p{0.4\linewidth}|p{0.02\linewidth}|p{0.03\linewidth}|}   
 \hline
\textbf{Sentiment} & \textbf{Description}  & \textbf{Examples} & \textbf{\textit{n}} & \textbf{\textit{p}}\\
\hline
 Positive & Positive descriptions of conversational agents (ECAs) or their features, highlighting favorable aspects or experiences.  & ``I felt very immersed and almost like I was talking to real people", ``Enjoyed the fact the AI characters used body language when speaking"& 33 & 0.53 \\
 \hline
Negative & Negative descriptions of conversational agents (ECAs) or their features, focusing on unfavorable aspects or experiences.& ``Delay in speech responses diminished sense of reality", ``They seemed a bit creepy. Talking to them felt like talking to a chatbot" & 12 & 0.19  \\
\hline
Mixed/Neutral & A combination of both positive and negative sentiments about conversational agents (ECAs), or expressions of neutral sentiment without a clear positive or negative bias. & ``I felt they were easy to converse with however it felt that they were a bit expressionless", ``It was interesting but easy to know they were AI" & 17 & 0.27 \\
\hline
  \end{tabular}
\end{table*}

\end{document}